\documentclass[aps,pre,reprint,floatfix]{revtex4-1}
\usepackage[english]{babel}
\usepackage{amsmath, amsthm, amssymb, latexsym, graphicx}
\usepackage[utf8]{inputenc}
\usepackage[usenames,dvipsnames]{xcolor}

\newcommand{\la}{\lambda}
\newcommand{\ten}[3]{#1^{#2}_{#3}(\textbf{r})}
\newcommand{\te}[3]{#1^{#2}_{#3}(-\textbf{r})}

\begin{document}

\title{von K\'arm\'an-Howarth equation for three-dimensional two-fluid plasmas}

\author{N. Andr\'es$^{1,2}$}
\author{P.D. Mininni$^{2,3}$}
\author{P. Dmitruk$^{2,3}$}
\author{D.O. G\'omez$^{1,2}$}
\affiliation{$^1$ Instituto de Astronom\'ia y F\'isica del Espacio, CONICET-UBA, 
                          Ciudad Universitaria, 1428, Buenos Aires, Argentina.\\
                  $^2$ Departamento de F\'{\i}sica, Facultad de Ciencias Exactas y 
                          Naturales, Universidad de Buenos Aires, Ciudad Universitaria, 1428,
                          Buenos Aires, Argentina.\\
                  $^3$ Instituto de F\'{\i}sica de Buenos Aires, CONICET-UBA, Ciudad
                          Universitaria, 1428, Buenos Aires, Argentina.}
\date{\today}

\begin{abstract}
We derive the von K\'arm\'an-Howarth equation for a full three dimensional incompressible two-fluid plasma. In the long-time limit and for very large Reynolds numbers we obtain the equivalent of the hydrodynamic ``four-fifth'' law. This exact law predicts the scaling of the third-order two-point correlation functions, and puts a strong constraint on the plasma turbulent dynamics. Finally, we derive a simple expression for the 4/5 law in terms of third-order structure functions, which is appropriate for comparison with in-situ measurements in the solar wind at different spatial ranges.
\end{abstract}

\maketitle

\section{Introduction}\label{intro}

In the paradigmatic case of incompressible hydrodynamic turbulence, \citet{vkh1938} obtained an evolution equation for the second-order correlation tensor under the assumptions of isotropy and homogeneity. This is the so-called von K\'arm\'an-Howarth (vKH) equation, and is one of the cornerstones of turbulence theory. The equation, which relates the time evolution of the second-order correlation velocity tensor to the divergence of the third-order correlation velocity tensor, has been extensively studied in the literature \citep{R1940,H1975,MY1975,F1995}. One of its most important corollaries is the so-called ``four-fifths'' law. Assuming the existence of an inertial energy range for very large Reynolds number, it predicts a linear scaling of the longitudinal two-point third-order velocity structure function with the distance between the two points. This scaling puts a strong constraint on the dynamics of fully developed turbulence. In particular, one crucial consequence is that the increment of the velocity field $\delta u$ between two points separated by $r$ is proportional to $r^{1/3}$, which in Fourier space leads to the famous Kolmogorov spectrum $E(k)\sim k^{-5/3}$ for the energy \citep{K1941a}.

Multiple attempts have been made to extend these results to turbulent plasmas. Chandrasekhar \cite{Ch1951} derived vKH equations in the one-fluid incompressible magnetohydrodynamic (MHD) approximation. Using Els\"{a}sser variables, \citet{P1998a,P1998b} derived the equations for third-order structure functions and for correlation functions, and assuming full isotropy (i.e., including mirror symmetry), homogeneity, and equipartition between kinetic and magnetic energy, they obtained the corresponding $4/5$ law. In the absence of mirror-symmetry the structure of the second-order correlation tensors is more complex. In particular, \citet{P2003} reported an exact equation for homogeneous and isotropic MHD turbulent flows with non-zero helicity. In the large Reynolds number and long-time limit, the authors recovered linear scaling for the third-order correlation tensors.

At spatial scales approaching the ion skin-depth (although still larger than the electron skin-depth) the MHD model is no longer appropriate. It can be replaced by the Hall-MHD (HMHD) model, which is an extension to MHD including the Hall current and the electron pressure (see, e.g., \cite{K1973}). The HMHD model has been extensively studied in recent years, both analytically \citep{T1983,MY1998,Mi2002,S2003} and numerically \citep{Mi2003,Ma2003,D2006}. \citet{Ga2008} derived the vKH equation for the three-dimensional (3D) incompressible HMHD equations in the non-helical case, i.e., when mirror-symmetry is considered, and obtained exact scaling laws for the third-order correlation tensors. {However, the HMHD model also has limitations to describe a plasma. First, tt ignores the inertia of the electrons, which plays a role in the nonlinear dynamics at small scales. Second, it misses kinetic plasma dissipation mechanisms such as Landau damping, cyclotron-resonant damping, and other mode-coupling dissipative processes. As a result, the HMHD model can only be expected to hold down to scales, of the order of the ion-skin depth.}

{Interestingly, many of the results for a turbulent flow described above can be derived independently of the mechanism of dissipation. Here we use this fact to} 
derive the vKH equation for the 3D incompressible two-fluid equations (vKH-TF) describing a fully ionized hydrogen plasma, extending all previous MHD and HMHD results, which can be regarded as particular cases in the proper asymptotic limits. As a result, we obtain scaling predictions for a turbulent plasma while retaining the whole dynamics of both the electron and ion flows throughout all the relevant spatial scales. {These predictions should be valid down to the largest scale in which a plasma dissipation mechanism is present. The derivation is done considering in particular applications in space plasmas. We start by considering some of the implications of the ``four-fifths'' law for turbulence in hydrodynamics and in MHD, followed by the derivation of the vKH-TF equation, and its relevance to the description of turbulent space plasmas.}

\section{The importance of the 4/5 law}\label{4/5lawHD}

{As mentioned in the Introduction, the vKH equation \citep{vkh1938} can be used as the starting point to obtain several essential results in turbulence theory. In particular, the hydrodynamic 4/5 law \citep{K1941b} states that in the limit of infinite Reynolds number, the longitudinal two-point third-order velocity structure function,
\begin{equation}
B_{\parallel\parallel\parallel}^{vvv}(\textbf{r}) =
    \left<\{[\textbf{v}(\textbf{x}+\textbf{r})-
    \textbf{v}(\textbf{x})]\cdot\textbf{r}\}^3\right> ,
\label{eq:struc}
\end{equation}
evaluated for spatial increments \textbf{r}, is given in terms of the spatial increment itself and of the mean energy dissipation per unit of mass $\varepsilon$ by
\begin{equation}\label{4/5hd}
B_{\parallel\parallel\parallel}^{vvv}(\textbf{r}) = -\frac{4}{5}\varepsilon r ,
\end{equation}
where $r=|\textbf{r}|$, and the angular bracket in Eq.~(\ref{eq:struc}) denotes ensemble average. The prefactor $4/5$ in Eq.~(\ref{eq:struc}) gives its name to this exact result.}

{It is interesting that this result is independent of the dissipation mechanisms acting in the flow, and it only requires the existence of some dissipation at sufficiently small scales to get rid off all the power injected at large scales. As such, this law links information that is accessible from flow measurements at intermediate scales (the third-order moment of the velocity structure function, associated to the energy flux transported among scales by non-linear effects), to small scale quantities such as the rate of energy dissipation. As a result,  this exact law has been used to identify the range of scales in which dissipation is negligible (the so-called ``inertial range'' of scales in a turbulent flow, for which the exact law holds), to meassure energy dissipation rates, and to define a Reynolds number without precise knowledge of the flow viscosity or of the microscopic dissipation mechanisms (see, e.g., \cite{WEY2007} and references therein).}

{The generalization of Eq.~\eqref{4/5hd} to the MHD case \citep{P1998a,P1998b} relates longitudinal two-point third-order structure functions of the velocity and magnetic fields, to the spatial increment and to the energy dissipation rate. Although the prefactor changes, the law is still known as ``four-fifths''. The validity of the MHD law has been subjected to several numerical testings (see, e.g., \citep{MI2009F,Bo2009,W2010,Y2012}). Among several important results, the 4/5 law in MHD has been used to measure the energy cascade rate at large scales of the solar wind \citep{SV2007,Sa2008,M2008}, to estimate the Reynolds number of solar wind turbulence \cite{WEY2007}, to investigate large-scale solar wind models \citep{M1999,MB2008}, to predict the decay of MHD turbulence \citep{Da2009,W2012}, and to determine scaling exponents in measurements and in simulations using the so-called Extended Self-Similarity (ESS) hypothesis \citep{R1993,G1997}. This last example highlights the importance of having an exact law: the ESS method uses the 4/5 law to calibrate structure functions, decreasing uncertainties and improving the analysis of the measurements.}

{The generalization of the vKH and the corresponding 4/5 law for a plasma would thus be useful not only to study scaling laws at large and intermediate turbulent plasma scales, but also to discern at what scale dissipation becomes relevant, and therefore the scaling would become invalid. In the solar wind and in space plasmas, there is a debate on whether scaling laws observed at high frequencies correspond to the inertial or dissipative ranges (see, e.g., \cite{M2011}). A derivation of exact laws for two-fluid equations for a fully ionized plasma is therefore a decisive step to elucidate this point, which should includes previous MHD and HMHD results as asymptotic limits. Next, we define the correlation tensors used to derive the 3D vKH-TF equation, and the two-fluid equations used to describe the plasma, and proceed to derive the 3D vKH-TF equation and the $4/5$ law.}

\section{Correlation tensors}\label{def} 

The properties and structure of second-order correlation tensors are extensively discussed in \cite{ORM1997}. A derivation of the general form of homogeneous third-order correlation tensors can be found in \cite{P2003}. For completeness, we briefly present the main results of tensorial algebra needed to obtain the vKH-TF equation. Given two solenoidal vector fields \textbf{a}(\textbf{x}) and \textbf{b}(\textbf{x}), we define the second-order correlation tensor as
\begin{align}
{R}_{ij}^{ab}(\textbf{x},\textbf{x}')=\left<a_i(\textbf{x})b_j(\textbf{x}')\right> = \left<a_i b_j' \right>,
\end{align}
where \textbf{x}'$=$\textbf{x}+\textbf{r}, and the angular bracket denotes ensemble average. Spatial homogeneity implies that all regions of space are similar so far as statistical properties are concerned, which suggests that the results of averaging over a large number of realizations can be obtained equally well by averaging over a large region of space for one realization \citep{Ba1953}. Therefore, under this assumption, the second-order correlation tensors depend only on the relative separation \textbf{r} and is independent of its origin.

The most general expression for a solenoidal second-order correlation tensor is \citep{Ba1953}
\begin{align}
\frac{\ten{R}{ab}{ij}}{c_{ab}}=f^{ab}(r)\delta_{ij}+\frac{r}{2}\partial_rf^{ab}(r)P_{ij}+\epsilon_{ijk}\frac{r_k}{r}\tilde{f}^{ab}(r) ,
\end{align}
where $c_{ab}=\bar{a}\bar{b}$, $\bar{a}$ is the root mean square value of \textbf{a}, and $c_{ab}f^{ab}(r)$ is the (dimensionless) longitudinal (along \textbf{r}) scalar correlation function. In addition, we have introduced a new pseudo-scalar function $\tilde{f}^{ab}(r)$ which is related to the not mirror-symmetric (or helical) part of the tensor. This pseudo-scalar function is, in general, not zero and related to the helicity of the fields involved. Finally, $P_{ij}(r)=\delta_{ij}- r_ir_j/r^2$ is the projector into the subspace of incompressible flows.

Third-order correlation tensors needed to derive the vKH-TF equation are of the form \begin{align}
S_{ikj}^{abc}(\textbf{x},\textbf{x}')=\left<a_i(\textbf{x})b_k(\textbf{x})c_j(\textbf{x}'))\right> = \left< a_i b_k c_j' \right> ,
\end{align}
where \textbf{c}(\textbf{x}) is another vector field. These tensors satisfy
\begin{eqnarray}
S_{ikj}^{abc}(\textbf{r}) & = & s_{11}(r)r_j\delta_{ik}+s_{12}(r)r_k\delta_{ij}+s_{13}(r)r_i\delta_{kj} \nonumber \\
&&+ s_3(r)r_ir_kr_j + \tilde{s}_{23}(r)r_i\epsilon_{kjm}r_m \nonumber \\ &&+ \tilde{s}_{22}(r)r_k\epsilon_{jim}r_m + \tilde{s}_{21}(r)r_j\epsilon_{ikm}r_m ,
\end{eqnarray}
where the seven scalar functions are even functions of $r$, as for the second-order correlation tensor. Note that we omitted the superscript $abc$ for the sake of simplicity. The incompressibility condition leads to relationships between these seven functions and reduces the problem to only four generating functions (see \cite{P2003}). If the first two indices (i.e., $i$ and $k$) are symmetric, the tensor is generated only by one scalar and one pseudo-scalar function (the latter related to helicity \citep{Ch1996,G2000,K2003}).

Finally, the second and third order structure functions are defined in terms of the increments of the vectors as
\begin{eqnarray}\label{sf2}
\ten{B}{ab}{ij} &=& \left<(a_i'-a_i)(b_j'-b_j)\right> , \\ \label{sf3}
\ten{B}{abc}{ikj} &=& \left<(a_i'-a_i)(b_k'-b_k)(c_j'-c_j)\right> .
\end{eqnarray}
Using homogeneity the following relations between structure functions and correlation tensors can be derived
\begin{eqnarray}\label{str}
\ten{B}{ab}{ij} &=& 2R_{ij}^{ab}(0) - \ten{R}{ab}{ij} - \te{R}{ab}{ij} , \\
\ten{B}{abc}{ijk} &=& \big[\ten{S}{abc}{ijk}-\te{S}{abc}{ijk}\big]\nonumber +\big[\ten{S}{acb}{ikj} \\ && -\te{S}{acb}{ikj}\big] + \big[\ten{S}{bca}{jki}-\te{S}{bca}{jki}\big] .
\end{eqnarray}

\section{Two fluid equations}\label{model}

The equations of motion for a quasi-neutral incompressible plasma of ions and electrons with mass $m_{i,e}$, charge $\pm e$, densities $n_{i}=n_{e}=n$, pressures $p^{(i,e)}$, and respective velocities $\textbf{v}$ and $\textbf{u}$ are \cite{A2014a,A2014b,A2016}
\begin{align}
m_en\frac{d \textbf{u}}{dt} &=  -en\left(\textbf{E}+\frac{1}{c}\textbf{u}\times\textbf{B}\right)-\boldsymbol\nabla p^{(e)} + \mu^{(e)} \nabla^2\textbf{u}- \textbf{D} , \label{eq:1} \\
m_in\frac{d \textbf{v}}{dt} &=  en\left(\textbf{E}+\frac{1}{c}\textbf{v}\times\textbf{B}\right)-\boldsymbol\nabla p^{(i)} + \mu^{(i)} \nabla^2\textbf{v}+ \textbf{D} , \label{eq:2} \\
\textbf{J} &= \frac{c}{4\pi}\boldsymbol\nabla\times\textbf{B}={en}(\textbf{v}-\textbf{u}) .
\end{align}
Here $d/dt = \partial/\partial t + {\bf u} \cdot \nabla$ is the total derivative, $\textbf{B}$ and $\textbf{E}$ are the magnetic and electric fields, $\textbf{J}$ is the electric current density, $c$ is the speed of light, $\mu^{(i,e)}$ are viscosities, and $\textbf{D}$ is the rate of momentum gained by ions due to collisions with electrons, and assumed proportional to the relative speed between species, $\textbf{D} = - nm_i\nu_{ie}(\textbf{v} - \textbf{u})$, where $\nu_{ie}$ is the collisional frequency of ions against electrons. Incompressibility implies $\boldsymbol\nabla\cdot\textbf{u} = 0 = \boldsymbol\nabla\cdot\textbf{v}$. {Note that these equations neglect kinetic plasma dissipation mechanisms due to either electrons or ions. As mentioned in Sec.~\ref{4/5lawHD}, the 4/5 law will hold for scales larger than any of the scales associated with these dissipation mechanisms, and as such, it should be independent of the actual damping mechanism acting in the plasma. The vKH-TF equation, on the other hand, will have explicit dissipation terms, and these terms may differ if other damping mechanisms are considered.}

Equations (\ref{eq:1}) and (\ref{eq:2}) can be written in dimensionless form in terms of a typical length $L_0$, the particle density $n$, a typical velocity $v_A=B_0/(4\pi nM)^{1/2}$ (the Alfv\'en velocity, where $B_0$ is a typical value of ${\bf B}$, and $M\equiv m_i+m_e$), and with the electric field in units of $E_0 = v_AB_0/c$,
\begin{align}\label{dlesse}
 \mu\frac{d \textbf{u}}{dt} &= -\frac{1}{\lambda}(\textbf{E}+\textbf{u}\times\textbf{B})-\boldsymbol\nabla p^{(e)} +\nu^{(e)}\nabla^2\textbf{u}-\frac{\textbf{d}}{\lambda} , \\
\label{dlessp}
 (1-\mu)\frac{d \textbf{v}}{dt} &= \frac{1}{\lambda}(\textbf{E}+\textbf{v}\times\textbf{B})-\boldsymbol\nabla p^{(i)}+\nu^{(i)}\nabla^2\textbf{v}+\frac{\textbf{d}}{\lambda} , \\  
\label{dlesso}
\textbf{J} &= \frac{1}{\lambda}(\textbf{v}-\textbf{u}) .
\end{align} 
where we have introduced the dimensionless parameters $\mu\equiv m_e/M$ and $\lambda\equiv c/\omega_{M}L_0$, where $\omega_{M}=(4\pi e^2n/M)^{1/2}$  has the form of a plasma frequency for a particle of mass $M$. The dimensionless momentum exchange rate is $\textbf{d} = -\eta\textbf{J}$, where $\eta = m_ic^2\nu_{ie}/(4\pi e^2nv_A L_0)$ is the (dimensionless) electric resistivity. Dimensionless ion and electron inertial lengths can be defined in terms of their corresponding plasma frequencies $\omega_{i,e}=(4\pi e^2n/m_{i,e})^{1/2}$ simply as $\lambda_{i,e}\equiv c/\omega_{i,e}L_0$, and their expressions in terms of $\mu$ and $\lambda$ are simply $\lambda_i=(1-\mu)^{1/2}\lambda$ and $\lambda_e=\mu^{1/2}\lambda$. Note that in the limit of electron inertia equal to zero, we obtain $\omega_{M}=\omega_{i}$, and therefore $\lambda = \lambda_i = c/\omega_{i}L_0$ reduces to the usual Hall parameter.

To obtain a hydrodynamic description of the two-fluid plasma, we can write \textbf{u} and \textbf{v} in terms of two vector fields (see \cite{A2014b}): the hydrodynamic velocity $\textbf{U} = (1-\mu)\textbf{v}+\mu\textbf{u}$, and $\textbf{J}$ as given by Eq.~\eqref{dlesso}. From these two fields, it is trivial to obtain \textbf{u} and \textbf{v} as $\textbf{u}=\textbf{U} - (1-\mu)\lambda\textbf{J}$ and $\textbf{v}=\textbf{U} + \mu\lambda\textbf{J}$. We will now see that this hydrodynamic description is useful to obtain the vKH-TF equation.  

\section{Results for a two-fluid plasma}\label{results}

\subsection{The von K\'arm\'an-Howarth equation}\label{3dvkh}

The $i$-th component of eqs.~\eqref{dlesse} and \eqref{dlessp} evaluated at point \textbf{x} is
\begin{eqnarray}\label{motione_i}
{} && \mu \partial_tu_i = - \mu u_k\partial_k u_i - \frac{1}{\la}\big[E_i+\epsilon_{ilm}U_lB_m  \nonumber \\ {} && -(1-\mu)\la\epsilon_{ilm}J_lB_m\big] - \partial_i p^{(e)} + \nu^{(e)}\partial_{kk}^2u_i + \frac{\eta}{\la}J_i  , \\ \label{motionp_i}
{} && (1-\mu) \partial_tv_i = - (1 - \mu) v_k\partial_{r_{k}} v_i + \frac{1}{\la}\big[E_i+\epsilon_{ilm}U_lB_m \nonumber \\ {} && + \mu\la\epsilon_{ilm}J_lB_m\big] -  \partial_i p^{(i)} + \nu^{(i)}\partial_{kk}^2v_i - \frac{\eta}{\la}J_i .
\end{eqnarray}

The time evolution equation for the second-order correlation tensor $\ten{R}{uu}{ij}$ results from Eq.~\eqref{motione_i} after multiplying by $u_j'= u_j(\textbf{x}+\textbf{r})$, and adding the $j$-th component of Eq.~\eqref{dlesse} at point $\textbf{x}'$ multiplied by $u_i$. The time evolution of $\ten{R}{vv}{ij}$ is obtained by performing similar operations on Eq.~\eqref{motionp_i}. The end result is
\begin{eqnarray}\label{motione_ij}
\partial_t\big[ &\mu& \ten{R}{uu}{ij} \big] =  \partial_{r_{k}}\big\{\mu\big[\ten{S}{uuu}{ikj}-\te{S}{uuu}{jki} \big] \nonumber \\
{} &-& (1-\mu)\big[\ten{S}{BBu}{ikj}-\te{S}{BBu}{jki} \big]\big\} - \big[\ten{R}{Eu}{ij} \nonumber \\
{} &+&\ten{R}{uE}{ij}\big]/\la \nonumber - \big[\epsilon_{ilm}\ten{S}{UBu}{lmj} + \epsilon_{jlm}\te{S}{UBu}{lmi}\big]/\la \nonumber \\
{} &+& 2\partial^2_{r_{k}r_{k}}\big[\nu^{(e)}\ten{R}{uu}{ij}\big]+\eta\big[\ten{R}{Ju}{ij}+\ten{R}{uJ}{ij}\big]/\la \\
\label{motionp_ij}
\partial_t\big[ ( &1& -\mu)\ten{R}{vv}{ij} \big] = \partial_{r_{k}}\big\{(1-\mu)\big[\ten{S}{vvv}{ikj}-\te{S}{vvv}{jki} \big]   \nonumber \\
{} &-& \mu\big[\ten{S}{BBv}{ikj}-\te{S}{BBv}{jki} \big]\big\} + \big[\ten{R}{Ev}{ij} +\ten{R}{uE}{ij}\big]/\la \nonumber \\
{} &+& \big[\epsilon_{ilm}\ten{S}{UBv}{lmj} + \epsilon_{jlm}\te{S}{UBv}{lmi}\big]/\la \nonumber \\
{} &+& 2\partial^2_{r_{k}r_{k}}\big[\nu^{(e)}\ten{R}{vv}{ij}\big]-\eta\big[\ten{R}{Jv}{ij}+\ten{R}{vJ}{ij}\big]/\la
\end{eqnarray}
where we have used the divergence-free condition for the fields $\textbf{u},\textbf{v}$, and $\textbf{B}$, $\partial_{r_k} \left< . \right>=\partial_k' \left< . \right>=-\partial_k \left< . \right>$ from homogeneity, and the relation 
\begin{align}
\ten{S}{abc}{ijk}=\left<a_i(\textbf{x})b_j(\textbf{x})c_k(\textbf{x}+\textbf{r})\right>=\nonumber\\\left<a_i(\textbf{x}+\textbf{r})b_j(\textbf{x}+\textbf{r})c_k(\textbf{x})\right>=\te{S}{abc}{ijk} .
\end{align}
Note that the gradient terms vanish because of isotropy \citep{Ba1953}. Adding eqs.~\eqref{motione_ij} and \eqref{motionp_ij} we obtain
\begin{eqnarray}\label{genera}
{} \partial_t \big[ &\mu& \ten{R}{uu}{ij} + (1-\mu)\ten{R}{vv}{ij} \big] - \big[\ten{R}{EJ}{ij}+\ten{R}{JE}{ij}\big] \nonumber \\
{} &=& \partial_{r_{k}}\big\{\mu\big[\ten{S}{uuu}{ikj}-\te{S}{uuu}{jki} \big] + (1-\mu)\big[\ten{S}{vvv}{ikj} \nonumber \\
{} &-& \te{S}{vvv}{jki} \big] 
-(1-\mu)\big[\ten{S}{BBu}{ikj}-\te{S}{BBu}{jki} \big] \nonumber \\
{} &-& \mu\big[\ten{S}{BBv}{ikj}-\te{S}{BBv}{jki} \big]\big\} + \big[\epsilon_{ilm}\ten{S}{UBJ}{lmj} \nonumber \\
{} &+& \epsilon_{jlm}\te{S}{UBJ}{lmi}\big] + 2\partial^2_{r_{k}r_{k}}\big[\nu^{(e)}\ten{R}{uu}{ij}+ \nu^{(i)}\ten{R}{vv}{ij}\big] \nonumber \\
{} &-& 2 \eta\ten{R}{JJ}{ij}.
\end{eqnarray}
The pseudo-tensor $\ten{S}{UBJ}{lmi}$ on the r.h.s~of Eq.~\eqref{genera} can be expressed in terms of the derivative of the proper tensor $\ten{S}{UBB}{lmi}$, since
\begin{align}
\epsilon_{ilm}\ten{S}{UBJ}{jmi} =  \epsilon_{ilm}\epsilon_{jpq}\partial_{r_p}\ten{S}{UBB}{lmq} = \partial_{r_p}\ten{T}{UBB}{ipj} .
\end{align}
We can write this tensor using the velocity of the species
\begin{equation}
\ten{T}{UBB}{ipj} = \mu\ten{T}{uBB}{ipj} + (1-\mu)\ten{S}{vBB}{ipj}.
\end{equation}
Finally, noting that $\ten{R}{JJ}{ij} = -\epsilon_{ipq}\epsilon_{jrs}\partial^2_{r_{p}r_{r}}\ten{R}{BB}{qs}$, Eq.~\eqref{genera} can be written as
\begin{eqnarray}\label{general}
\partial_t\big[ &\mu& \ten{R}{uu}{ij} + (1-\mu)\ten{R}{vv}{ij}\big] - \big[\ten{R}{EJ}{ij}+\ten{R}{JE}{ij}\big] \nonumber \\
{} &=& \partial_{r_{k}}\big\{\mu\big[\ten{S}{uuu}{ikj}-\te{S}{uuu}{jki} \big] + (1-\mu)\big[\ten{S}{vvv}{ikj} \nonumber \\
{} &-& \te{S}{vvv}{jki} \big] -(1-\mu)\big[\ten{S}{BBu}{ikj}-\te{S}{BBu}{jki} \big] \nonumber \\
{} &-& \mu\big[\ten{S}{BBv}{ikj}-\te{S}{BBv}{jki} \big] + \mu\big[\ten{T}{uBB}{ikj}\nonumber \\
{} &+& \te{T}{uBB}{ikj}\big]+(1-\mu)\big[\ten{T}{vBB}{ikj}+\te{T}{vBB}{ikj}\big]\big\} \nonumber \\
{} &+& 2\partial^2_{r_{k}r_{k}}\big[\nu^{(e)}\ten{R}{uu}{ij}+\nu^{(i)}\ten{R}{vv}{ij} \big] + \nonumber \\ &+& 2\eta\epsilon_{ipq}\epsilon_{jrs}\partial^2_{r_{p}r_{r}}\ten{R}{BB}{qs}.
\end{eqnarray}
Equation \eqref{general} is an exact law, valid even for anisotropic turbulence \citep{M1981,P2008,C2009}. This is our first main result.

We can now take the trace of Eq.~\eqref{general},
\begin{eqnarray}\label{vkh_parallel1}
\partial_t\big[ &\mu& \ten{R}{uu}{ii} + (1-\mu)\ten{R}{vv}{ii} + \ten{R}{BB}{ii} \big] \nonumber \\
{} &=& \partial_{r_{k}}\big\{2\mu\ten{S}{uuu}{iki} + 2(1-\mu)\ten{S}{vvv}{iki} \nonumber \\ {} &-& 2(1-\mu)\ten{S}{BBu}{iki} -2\mu\ten{S}{BBv}{iki} + \mu\big[\ten{T}{uBB}{iki} \nonumber \\
{} &+& \te{T}{uBB}{iki}\big]+(1-\mu)\big[\ten{T}{vBB}{iki}+\te{T}{vBB}{iki}\big]\big\} \nonumber \\
{} &+& 2\partial^2_{r_{k}r_{k}}\big[\nu^{(e)}\ten{R}{uu}{ii}+\nu^{(i)}\ten{R}{vv}{ii}+\eta\ten{R}{BB}{ii}\big],
\end{eqnarray}
where, for $a$ equal to $u$ or $v$,
\begin{eqnarray}
\ten{T}{aBB}{iki} &=&\epsilon_{ilm}\epsilon_{ikq}\ten{S}{aBB}{lmq} = (\delta_{lk}\delta_{mq}-\delta_{lq}\delta_{mk})\ten{S}{aBB}{lmq} \nonumber \\
{} &=& \ten{S}{aBB}{kqq}-\ten{S}{aBB}{qkq}.
\end{eqnarray}
From homogeneity, we have noted that
\begin{eqnarray}\label{mag}
\ten{R}{EJ}{ii}+\ten{R}{JE}{ii} &=& -\partial_t\ten{R}{BB}{ii}.
\end{eqnarray}
In the asymptotic limit of $\textbf r \rightarrow 0$, Eq.~\eqref{mag} corresponds to twice the time variation of the total magnetic energy.

Now, assuming isotropy, we introduce the explicit form of the second- and third-order correlation tensors,
\begin{eqnarray}
{} && \frac{\ten{R}{aa}{ii}}{c_{aa}} = (3+r\partial_r)f^{aa}(r) , \\
{} && \frac{\ten{S}{aab}{iki}}{c_{aab}} = \frac{4k^{aab} +r\partial_rk^{aab}}{2r}r_k= \text{S}^{aab}(r)r_k ,
\end{eqnarray}
\begin{eqnarray}
\frac{\ten{S}{abb}{kii}-\ten{S}{abb}{iki}}{c_{abb}} &=& \frac{2k^{abb}+r\partial_rk^{abb}-4q^{abb}}{r}r_k  \nonumber \\
{} &=& \text{T}^{abb}(r)r_k ,
\end{eqnarray}
where
\begin{eqnarray}
f^{aa}(r) &=& R_{\parallel\parallel}^{aa}/c_{aa} , \,\,\,\,
k^{aab}(r) = \ten{S}{aab}{\parallel\parallel\parallel}/c_{aab} , \nonumber \\
q^{abb}(r) &=& \ten{S}{abb}{\perp\parallel\perp}/c_{abb} , \nonumber
\end{eqnarray}
are the longitudinal and transversal correlations of the fields $a,b=u,v$ and $B$, and where $c_{aa}=\bar{a}\bar{a}$ and $c_{abb}=\bar{a}\bar{b}\bar{b}$. By longitudinal ($\parallel$) and transversal ($\perp$), we mean in the direction along and perpendicular to the displacement vector \textbf{r}, respectively. Using these expressions and after some manipulation, Eq.~\eqref{vkh_parallel1} reduces to
\begin{eqnarray}\label{vkh_parallel2}
( 3 &+& r\partial_r)\partial_t\big[\mu c_{uu}f^{uu} + (1-\mu) c_{vv}f^{vv} + c_{BB}f^{BB}\big] \nonumber \\
{} &=& 2(3+r\partial_r)\big\{\mu c_{uuu} S^{uuu} + (1-\mu) c_{vvv} S^{vvv} \nonumber \\
{} &-& (1-\mu) c_{BBu} S^{BBu} - \mu c_{BBv} S^{BBv} + \mu c_{uBB} T^{uBB} \nonumber \\
{} &+& (1-\mu) T^{vBB} 
+2\partial_r\big[r^4\partial_r \big(\nu^{(e)}c_{uu}f^{uu}+\nu^{(i)}c_{vv}f^{vv} \nonumber \\
{} &+& \eta c_{BB}f^{BB}\big)\big]/r^4\big\}
\end{eqnarray}
where we have used identities for isotropic turbulence \citep{Ba1953}
\begin{align}
\partial_{kk} &= \partial^2_{rr} + \frac{2}{r}\partial_r ,\\
\big(\partial_{rr}^2+\frac{2}{r}\partial_r\big)(3+r\partial_r) &= (3+r\partial_r)\frac{1}{r^4}\partial_r(r^4\partial_r).
\end{align}
A first integral of Eq.~\eqref{vkh_parallel2} is
\begin{eqnarray}\label{vkh}
\partial_t\big[&\mu& c_{uu}f^{uu} + (1-\mu) c_{vv}f^{vv} + c_{BB}f^{BB} \big]\nonumber \\
{} &=& 2\big[\mu c_{uuu} S^{uuu} + (1-\mu) c_{vvv} S^{vvv} \nonumber \\
{} &-& (1-\mu) c_{BBu} S^{BBu} - \mu c_{BBv} S^{BBv} + \mu c_{uBB} T^{uBB} \nonumber \\
{} &+& (1-\mu) T^{vBB} \big] +2\partial_r\big[r^4\partial_r \big(\nu^{(e)}c_{uu}f^{uu} \nonumber \\
{} &+& \nu^{(i)}c_{vv}f^{vv}+\eta c_{BB}f^{BB})\big]/r^4.
\end{eqnarray}
This exact relation is the von K\'arm\'an-Howarth equation for an incompressible two-fluid plasma, which is the second main result of the present paper.

\subsection{The 4/5 law}\label{4/5law}

The vKH-TF Eq.~\eqref{vkh} can be writen in terms of the structure functions using the relation
\begin{align}
\ten{R}{aa}{\parallel\parallel}=\left<a_{\parallel}a_{\parallel}\right>-\ten{B}{aa}{\parallel\parallel}/2,
\end{align}
where we have used Eq.~\eqref{str}. Therefore
\begin{eqnarray}
\partial_t\bigg[ &\mu& \left<u^2_{\parallel}\right>+(1-\mu)\left< v^2_{\parallel}\right>+\left<B^2_{\parallel}\right>\bigg] \nonumber \\
{} &-& \partial_t\big[\mu B^{uu}_{\parallel\parallel}+(1-\mu)B^{vv}_{\parallel\parallel}+B^{BB}_{\parallel\parallel}\big] \nonumber \\
{} &=& 2\big[ \mu c_{uuu} S^{uuu} + (1-\mu) c_{vvv} S^{vvv} - (1-\mu) c_{BBu} S^{BBu} \nonumber \\
{} &-& \mu c_{BBv} S^{BBv} + \mu c_{uBB} T^{uBB} + (1-\mu) T^{vBB}\big] \nonumber \\
{} &+& 2\nu^{(e)} r^{-4} \partial_r\left[r^4\partial_r\left(\left<u^2_{\parallel}\right>-B_{\parallel\parallel}^{uu}/2\right)\right] \nonumber \\
{} &+& 2\nu^{(i)} r^{-4}\partial_r\left[r^4\partial_r\left(\left<v^2_{\parallel}\right>-B_{\parallel\parallel}^{vv}/2\right)\right]\nonumber \\
{} &+&  2\eta r^{-4}\partial_r\left[r^4\partial_r\left(\left<B^2_{\parallel}\right>-B_{\parallel\parallel}^{BB}/2\right) \right] ,
\end{eqnarray}
where we identify the mean energy dissipation rate per unit mass $\varepsilon_T$ for isotropic turbulence as
\begin{equation}
\partial_t\left(\mu\left<u^2_{\parallel}\right>+(1-\mu)\left< v^2_{\parallel}\right>+\left<B^2_{\parallel}\right>\right) = -\frac{2}{3}\varepsilon_T.
\end{equation} 

Hereafter we adopt the usual assumption for fully developed turbulence \citep{F1995}, i.e., we use the long-time limit in which a stationary regime is reached for sufficiently large Reynolds numbers, and in which $\varepsilon_T$ tends to a finite positive value. Therefore, at the inertial range
\begin{align}\label{pre4/5}
-\frac{2}{3}r\varepsilon_T &= \mu c_{uuu} (4k^{uuu} +r\partial_rk^{uuu}) \nonumber \\
{} &+ (1-\mu) c_{vvv} (4k^{vvv} +r\partial_rk^{vvv}) \nonumber \\
{} &- (1-\mu) c_{BBu} (4k^{BBu} +r\partial_rk^{BBu}) \nonumber \\
{} &- \mu c_{BBv} (4k^{BBv} +r\partial_rk^{BBv}) \nonumber \\
{} &+ 8\mu c_{uBB} (k^{uBB}+r\partial_rk^{uBB}/2-2q^{uBB}) \nonumber \\
{} &+ 8(1-\mu)c_{vBB} (k^{vBB}+r\partial_rk^{vBB}/2-2q^{vBB}).
\end{align}
Equation \eqref{pre4/5} can also be written in terms of the third-order correlation tensors as
\begin{align}\label{almost4/5}
-\frac{1}{6}r\varepsilon_T =\mu(S_{\parallel\perp\perp}^{uuu}+\frac{1}{2}S_{\parallel\parallel\parallel}^{uuu}) + (1-\mu) (S_{\parallel\perp\perp}^{vvv}+\frac{1}{2}S_{\parallel\parallel\parallel}^{vvv}) \nonumber \\ - (1-\mu) (S_{\parallel\perp\perp}^{BBu}+\frac{1}{2}S_{\parallel\parallel\parallel}^{BBu}) - \mu (S_{\parallel\perp\perp}^{BBv}+\frac{1}{2}S_{\parallel\parallel\parallel}^{BBv}) \nonumber \\ + \mu (S_{\parallel\perp\perp}^{uBB}-S_{\perp\parallel\perp}^{uBB}) + (1-\mu) (S_{\parallel\perp\perp}^{vBB}-S_{\perp\parallel\perp}^{vBB})
\end{align}
where $\perp=2$ or 3 (no summation on $\perp$). Finally, using eqs.~\eqref{sf2}, \eqref{sf3} and the incompressibility condition ($\ten{S}{abb}{\perp\perp\parallel}=-\ten{B}{baa}{\parallel\parallel\parallel}/2$), we write Eq.~\eqref{almost4/5} as a function of the third-order structure functions
\begin{align}\label{4/5}
-\frac{4}{3}r\varepsilon_T = \bigg[\ten{B}{vvv}{\parallel ii}+\ten{B}{vBB}{\parallel ii}-\ten{B}{BvB}{\parallel ii}-\ten{B}{BuB}{\parallel ii} \bigg] \nonumber \\+ \mu\bigg[\ten{B}{uuu}{\parallel ii}+\ten{B}{uBB}{\parallel ii}-\ten{B}{vvv}{\parallel ii}-\ten{B}{vBB}{\parallel ii}\bigg].
\end{align}
This equation is the $4/5$ law for the two-fluid plasma, and is the third main result of the present paper. 

\subsection{Discussion}\label{disc}

Equations \eqref{general}, \eqref{vkh}, and \eqref{4/5} give respectively: (1) an exact relation for the correlation functions of anisotropic turbulence (e.g., in the presence of a guide field) in a two-species incompressible plasma, (2) the vKH equation for isotropic turbulence, and (3) the $4/5$ law for the scaling of the flux in the inertial range. At the large scales, i.e., when $\lambda\rightarrow0$ and $\mu\rightarrow0$, we recover the MHD results \citep{Ch1951,P1998a,P1998b}, and the hydrodynamic result when the magnetic field is taken equal to zero. When only $\mu\rightarrow0$ we obtain the HMHD case, previously studied by \citet{Ga2008}. However, unlike the results in \citet{Ga2008}, here we also express the results in terms of structure functions. {Also, it is important to emphasize that our expressions are written in terms of the velocity of each species in the plasma, which allow easier comparison with observations in space physics and with data from numerical simulations}.

The exact Eq.~\eqref{4/5} implies a scaling law for the third-order structure functions in the large Reynolds numbers and long-time limits. This law imposes correlations between the basic fields \textbf{u}, \textbf{v} and \textbf{B}, putting a strong constraint on the plasma turbulent dynamics. Finally, the equivalent expression involving only third-order structure functions is appropriate for comparisons with {\it in-situ} measurements in the solar wind at different spatial ranges (such as those in \cite{S2009}), or with laboratory plasmas.

Under certain assumptions, eqs.~\eqref{vkh} and \eqref{4/5} also provide predictions for the scaling of the energy spectrum in a turbulent plasma. At the largest scales, using Eq.~\eqref{4/5} and assuming energy equipartition, we recover the well known result $\delta B\sim r^{1/3}$, which corresponds to the Kolmogorov spectrum for the total (kinetic plus magnetic) energy $E(k)\sim k^{-5/3}$. At intermediate scales (assuming $\mu=0$ and $r>\lambda$), and using the fact that the electron velocity $\delta u$ is proportional to $\delta B/r$, Eq.~\eqref{4/5} leads to $\delta B\sim r^{2/3}$, which corresponds to a magnetic energy spectrum $E_B(k)\sim k^{-7/3}$. Finally, at the smallest scales ($r<\lambda_e$) where the terms proportional to $\mu$ in Eq.~\eqref{4/5} become dominant, a new scaling for the energy inertial range emerges. At these scales, using $\delta u\sim \delta B/r$, we obtain $\delta B\sim r^{4/3}$, and therefore $E_{B}(k)\sim k^{-11/3}$. This scaling was recently obtained from dimensional arguments and observed in numerical simulations in \citep{A2014b}. It is also remarkable that these three different scalings are compatible with previous theoretical and numerical results \citep{M2010,G2008,Mi2007,Mi2003,Bi1997,MG2010} and with solar wind observations \citep{SV2007,MB2008,S2009,S2010,S2011,A2009}.

Regarding the intermediate HMHD range, it is worth mentioning that an energy spectrum with a slope equal to $-7/3$ is incompatible with the assumption of an asymptotic separation of scales between the forced scales and the dissipative (assumed small) scales, as for a spectrum steeper than $-2$ dissipation peaks at large scales (i.e., in the HMHD range, at wavenumber $k\sim 1/\lambda$). However, at scales below the electron skin depth (where the HMHD model is no longer appropriate), the total energy spectrum (dominated by electron kinetic energy) is $E_k\sim k^{-5/3}$ \citep{A2014b}. Thus, as long as the HMHD range in the plasma is not too broad {(and as long as kinetic plasma damping mechanisms can be neglected at those scales)}, the scalings obtained above should hold. The dissipation anomaly for steep spectra has been studied before in the literature, for instance, in regularized MHD models \citep{P2006,H2002}.

\section{Conclusions}\label{conclus}

{We derived the von K\'arm\'an-Howarth equation, and its corresponding 4/5 law, for a 3D incompressible two-fluid plasma model. In particular, the derived vKH-TF equation can be written compactly in the usual form found for hydrodynamics \citep{R1940} as
\begin{equation}
\frac{\partial Q}{\partial t} = - \frac{\partial T}{\partial r} +
    \frac{\partial}{\partial r} \left[ \frac{1}{r^4} 
    \frac{\partial}{\partial r}r^4 D(Q)\right] ,
\end{equation}
where $Q$ is a function associated to the second-order correlation tensors of the fields, $T$ to the third-order correlation tensors, and $D(Q)$ takes into account all dissipative effects. The structure of the functions $Q$ and $T$ [see., e.g., Eq.~(\ref{general}) valid in the more general anisotropic case, and Eq.~(\ref{vkh}) valid in the isotropic case] is independent of the mechanism of dissipation acting on the plasma. However, the structure of the function $D(Q)$ depends on the dissipation mechanism, and should therefore be expected to change if kinetic plama dissipation mechanisms (e.g, Landau damping, cyclotron-resonant damping, or other mode-coupling dissipative processes) are present.}

{To obtain the generalization of the 4/5 law for a two-fluid plasma [Eq.~(\ref{4/5})] we adopted the usual assumption of fully developed turbulence, where an asymptotic regime is expected to be reached for sufficiently large Reynolds numbers. This allows for the existence of a range of scales for which dissipation mechanisms are negligible (i.e., $D\approx 0$ at those scales), and as a result $\partial_t Q$ can be associated to the mean energy dissipation rate per unit mass. Note that this is independent of the particular dissipation mechanisms in the plasma, as long as they act at sufficiently small scales, and as long as they allow for an asymptotic regime to be reached for sufficiently large scale separation. Another equivalent approach to derive the 4/5 law is to consider that in the turbulent steady regime, for which $\partial_tQ\approx0$, the dissipative term must be equal to the mean power injected into the system per unit mass, which is equal to $-\varepsilon_T$. Therefore, regardless of the particular mechanisms of dissipation present in the plasma (and in particular, in the solar wind), Eq.~(\ref{4/5}) provides an exact law which should hold as long as the energy injection rate balances the energy dissipation rate, and for all scales larger than the scale at which collisional or kinetic plasma dissipation mechanisms become dominant. The length of this scale will depend on specific properties of the plasma considered. As an example, in the solar wind at 1 AU dissipation mechanisms seem to become relevant for frequencies of $0.5$ Hz \cite{L1998}, although there is a debate on whether this scale is indeed dissipative or inertial \cite{M2011}.}

{The 4/5 law for a two-fluid plasma thus includes the effect of ion and electron inertia in the scaling of turbulence, and generalizes previous results obtained for MHD and HMHD. For scales larger than dissipative scales, it implies a specific scaling for structure functions of the velocity of each species and of the magnetic field, which can be related to the energy dissipation rate at the smallest scales. Therefore, it provides a way to test if a range of scales in a plasma is inertial or dissipative: if scaling laws observed in that range satisfy Eq.~(\ref{4/5}), then the scaling is the result of a turbulent cascade. It also provides a way to calibrate observations (e.g., higher order structure functions) against the 4/5 law, as often done when the ESS hypothesis are used to analyze data. In the range of scales in which it holds, it implies scaling laws for the energy spectrum at scales larger and smaller than the ion skin depth, as discussed in Sec.~\ref{disc}. Equation (\ref{4/5}) gives a way to quantify the total energy dissipation rate per unit mass in a plasma from observations at scales larger than the kinetic scales at which the dissipation mechanisms become dominant. Finally, since the expressions are given in terms of the velocity field of each species, the procedure used here can be extended to consider multi-species plasmas in a straightforward fashion.}

In the study of turbulent flows, exact laws provide an essential tool to analyze data and understand non-linear cascades. Over the last years, the sustained increase in the spatial and temporal resolution of space missions such as Cluster (ESA) or the new NASA MMS (Magnetospheric MultiScale) mission has opened the possibility to study a number of small-scale plasma phenomena as never before. The exact laws derived here allow investigation of the nature of turbulent magnetic field fluctuations at a broad range of scales in space plasmas, and will be indispensable to understand the nature of turbulence at the smallest scales in the solar wind.

\section*{Acknowledgments}

We acknowledge support from grants No.~PIP 11220090100825, UBACyT 20020110200359 and 20020100100315, and PICT 2011-0454 and 2011-1529.

\bibliographystyle{apsrev4-1}
\bibliography{cites}

\begin{thebibliography}{58}%
\makeatletter
\providecommand \@ifxundefined [1]{%
 \@ifx{#1\undefined}
}%
\providecommand \@ifnum [1]{%
 \ifnum #1\expandafter \@firstoftwo
 \else \expandafter \@secondoftwo
 \fi
}%
\providecommand \@ifx [1]{%
 \ifx #1\expandafter \@firstoftwo
 \else \expandafter \@secondoftwo
 \fi
}%
\providecommand \natexlab [1]{#1}%
\providecommand \enquote  [1]{``#1''}%
\providecommand \bibnamefont  [1]{#1}%
\providecommand \bibfnamefont [1]{#1}%
\providecommand \citenamefont [1]{#1}%
\providecommand \href@noop [0]{\@secondoftwo}%
\providecommand \href [0]{\begingroup \@sanitize@url \@href}%
\providecommand \@href[1]{\@@startlink{#1}\@@href}%
\providecommand \@@href[1]{\endgroup#1\@@endlink}%
\providecommand \@sanitize@url [0]{\catcode `\\12\catcode `\$12\catcode
  `\&12\catcode `\#12\catcode `\^12\catcode `\_12\catcode `\%12\relax}%
\providecommand \@@startlink[1]{}%
\providecommand \@@endlink[0]{}%
\providecommand \url  [0]{\begingroup\@sanitize@url \@url }%
\providecommand \@url [1]{\endgroup\@href {#1}{\urlprefix }}%
\providecommand \urlprefix  [0]{URL }%
\providecommand \Eprint [0]{\href }%
\providecommand \doibase [0]{http://dx.doi.org/}%
\providecommand \selectlanguage [0]{\@gobble}%
\providecommand \bibinfo  [0]{\@secondoftwo}%
\providecommand \bibfield  [0]{\@secondoftwo}%
\providecommand \translation [1]{[#1]}%
\providecommand \BibitemOpen [0]{}%
\providecommand \bibitemStop [0]{}%
\providecommand \bibitemNoStop [0]{.\EOS\space}%
\providecommand \EOS [0]{\spacefactor3000\relax}%
\providecommand \BibitemShut  [1]{\csname bibitem#1\endcsname}%
\let\auto@bib@innerbib\@empty
\bibitem [{\citenamefont {von K\'arm\'an}\ and\ \citenamefont
  {Howarth}(1938)}]{vkh1938}%
  \BibitemOpen
  \bibfield  {author} {\bibinfo {author} {\bibfnamefont {T.}~\bibnamefont {von
  K\'arm\'an}}\ and\ \bibinfo {author} {\bibfnamefont {L.}~\bibnamefont
  {Howarth}},\ }\href@noop {} {\bibfield  {journal} {\bibinfo  {journal} {Proc.
  R. Soc. London}\ }\textbf {\bibinfo {volume} {164}},\ \bibinfo {pages} {192}
  (\bibinfo {year} {1938})}\BibitemShut {NoStop}%
\bibitem [{\citenamefont {Robertson}(1940)}]{R1940}%
  \BibitemOpen
  \bibfield  {author} {\bibinfo {author} {\bibfnamefont {H.~P.}\ \bibnamefont
  {Robertson}},\ }\href@noop {} {\bibfield  {journal} {\bibinfo  {journal}
  {Proc Cambr. Phil. Soc.}\ }\textbf {\bibinfo {volume} {36}},\ \bibinfo
  {pages} {209} (\bibinfo {year} {1940})}\BibitemShut {NoStop}%
\bibitem [{\citenamefont {Hinze}(1975)}]{H1975}%
  \BibitemOpen
  \bibfield  {author} {\bibinfo {author} {\bibfnamefont {J.~O.}\ \bibnamefont
  {Hinze}},\ }\href@noop {} {\emph {\bibinfo {title} {Turbulence (2nd ed.)}}}\
  (\bibinfo  {publisher} {New York: McGraw-Hill},\ \bibinfo {year}
  {1975})\BibitemShut {NoStop}%
\bibitem [{\citenamefont {Monin}\ and\ \citenamefont {Yaglom}(1975)}]{MY1975}%
  \BibitemOpen
  \bibfield  {author} {\bibinfo {author} {\bibfnamefont {A.~S.}\ \bibnamefont
  {Monin}}\ and\ \bibinfo {author} {\bibfnamefont {A.~M.}\ \bibnamefont
  {Yaglom}},\ }\href@noop {} {\emph {\bibinfo {title} {Statistical Fluid
  Mechanics: Mechanics of Turbulence}}},\ Vol.~\bibinfo {volume} {2}\ (\bibinfo
   {publisher} {Cambridge, MA: MIT Press.},\ \bibinfo {year}
  {1975})\BibitemShut {NoStop}%
\bibitem [{\citenamefont {Frisch}(1995)}]{F1995}%
  \BibitemOpen
  \bibfield  {author} {\bibinfo {author} {\bibfnamefont {U.}~\bibnamefont
  {Frisch}},\ }\href@noop {} {\emph {\bibinfo {title} {Turbulence: The Legacy
  of A. N. Kolmogorov}}}\ (\bibinfo  {publisher} {Cambridge University
  Press.},\ \bibinfo {year} {1995})\BibitemShut {NoStop}%
\bibitem [{\citenamefont {Kolmogorov}(941a)}]{K1941a}%
  \BibitemOpen
  \bibfield  {author} {\bibinfo {author} {\bibfnamefont {A.~N.}\ \bibnamefont
  {Kolmogorov}},\ }\href@noop {} {\bibfield  {journal} {\bibinfo  {journal}
  {Akademiia Nauk SSSR Doklady}\ }\textbf {\bibinfo {volume} {30}},\ \bibinfo
  {pages} {301} (\bibinfo {year} {1941a})}\BibitemShut {NoStop}%
\bibitem [{\citenamefont {Chandrasekhar}(1951)}]{Ch1951}%
  \BibitemOpen
  \bibfield  {author} {\bibinfo {author} {\bibfnamefont {S.}~\bibnamefont
  {Chandrasekhar}},\ }\href@noop {} {\bibfield  {journal} {\bibinfo  {journal}
  {Proceedings of the Royal Society of London A: Mathematical, Physical and
  Engineering Sciences}\ }\textbf {\bibinfo {volume} {204}},\ \bibinfo {pages}
  {435} (\bibinfo {year} {1951})}\BibitemShut {NoStop}%
\bibitem [{\citenamefont {Politano}\ and\ \citenamefont
  {Pouquet}(998a)}]{P1998a}%
  \BibitemOpen
  \bibfield  {author} {\bibinfo {author} {\bibfnamefont {H.}~\bibnamefont
  {Politano}}\ and\ \bibinfo {author} {\bibfnamefont {A.}~\bibnamefont
  {Pouquet}},\ }\href@noop {} {\bibfield  {journal} {\bibinfo  {journal} {Phys.
  Rev. E}\ }\textbf {\bibinfo {volume} {57}},\ \bibinfo {pages} {R21} (\bibinfo
  {year} {1998a})}\BibitemShut {NoStop}%
\bibitem [{\citenamefont {Politano}\ and\ \citenamefont
  {Pouquet}(998b)}]{P1998b}%
  \BibitemOpen
  \bibfield  {author} {\bibinfo {author} {\bibfnamefont {H.}~\bibnamefont
  {Politano}}\ and\ \bibinfo {author} {\bibfnamefont {A.}~\bibnamefont
  {Pouquet}},\ }\href@noop {} {\bibfield  {journal} {\bibinfo  {journal}
  {Geophys. Res. Lett.}\ }\textbf {\bibinfo {volume} {25}},\ \bibinfo {pages}
  {273} (\bibinfo {year} {1998b})}\BibitemShut {NoStop}%
\bibitem [{\citenamefont {Politano}\ \emph {et~al.}(2003)\citenamefont
  {Politano}, \citenamefont {Gomez},\ and\ \citenamefont {Pouquet}}]{P2003}%
  \BibitemOpen
  \bibfield  {author} {\bibinfo {author} {\bibfnamefont {H.}~\bibnamefont
  {Politano}}, \bibinfo {author} {\bibfnamefont {T.}~\bibnamefont {Gomez}}, \
  and\ \bibinfo {author} {\bibfnamefont {A.}~\bibnamefont {Pouquet}},\
  }\href@noop {} {\bibfield  {journal} {\bibinfo  {journal} {Phys. Rev. E}\
  }\textbf {\bibinfo {volume} {68}},\ \bibinfo {pages} {026315} (\bibinfo
  {year} {2003})}\BibitemShut {NoStop}%
\bibitem [{\citenamefont {Krall}\ and\ \citenamefont
  {Trivelpiece}(1973)}]{K1973}%
  \BibitemOpen
  \bibfield  {author} {\bibinfo {author} {\bibfnamefont {N.~A.}\ \bibnamefont
  {Krall}}\ and\ \bibinfo {author} {\bibfnamefont {A.~W.}\ \bibnamefont
  {Trivelpiece}},\ }\href@noop {} {\bibfield  {journal} {\bibinfo  {journal}
  {in Principle of Plasma Physics}\ ,\ \bibinfo {pages} {89}} (\bibinfo {year}
  {1973})}\BibitemShut {NoStop}%
\bibitem [{\citenamefont {Turner}(1983)}]{T1983}%
  \BibitemOpen
  \bibfield  {author} {\bibinfo {author} {\bibfnamefont {L.}~\bibnamefont
  {Turner}},\ }\href@noop {} {\bibfield  {journal} {\bibinfo  {journal} {IEEE
  Trans. Plasma Sci.}\ }\textbf {\bibinfo {volume} {14}},\ \bibinfo {pages}
  {849} (\bibinfo {year} {1983})}\BibitemShut {NoStop}%
\bibitem [{\citenamefont {Mahajan}\ and\ \citenamefont
  {Yoshida}(1998)}]{MY1998}%
  \BibitemOpen
  \bibfield  {author} {\bibinfo {author} {\bibfnamefont {S.~M.}\ \bibnamefont
  {Mahajan}}\ and\ \bibinfo {author} {\bibfnamefont {Z.}~\bibnamefont
  {Yoshida}},\ }\href@noop {} {\bibfield  {journal} {\bibinfo  {journal} {Phys.
  Rev. Lett.}\ }\textbf {\bibinfo {volume} {81}},\ \bibinfo {pages} {4863}
  (\bibinfo {year} {1998})}\BibitemShut {NoStop}%
\bibitem [{\citenamefont {Mininni}\ \emph {et~al.}(2002)\citenamefont
  {Mininni}, \citenamefont {G\'omez},\ and\ \citenamefont {Mahajan}}]{Mi2002}%
  \BibitemOpen
  \bibfield  {author} {\bibinfo {author} {\bibfnamefont {P.~D.}\ \bibnamefont
  {Mininni}}, \bibinfo {author} {\bibfnamefont {D.~O.}\ \bibnamefont
  {G\'omez}}, \ and\ \bibinfo {author} {\bibfnamefont {S.~M.}\ \bibnamefont
  {Mahajan}},\ }\href@noop {} {\bibfield  {journal} {\bibinfo  {journal}
  {Astrophys. J.}\ }\textbf {\bibinfo {volume} {567}},\ \bibinfo {pages} {L81}
  (\bibinfo {year} {2002})}\BibitemShut {NoStop}%
\bibitem [{\citenamefont {Sahraoui}\ \emph {et~al.}(2003)\citenamefont
  {Sahraoui}, \citenamefont {Belmont},\ and\ \citenamefont {Rezeau}}]{S2003}%
  \BibitemOpen
  \bibfield  {author} {\bibinfo {author} {\bibfnamefont {F.}~\bibnamefont
  {Sahraoui}}, \bibinfo {author} {\bibfnamefont {G.}~\bibnamefont {Belmont}}, \
  and\ \bibinfo {author} {\bibfnamefont {L.}~\bibnamefont {Rezeau}},\
  }\href@noop {} {\bibfield  {journal} {\bibinfo  {journal} {Physics of
  Plasmas}\ }\textbf {\bibinfo {volume} {10}},\ \bibinfo {pages} {1325}
  (\bibinfo {year} {2003})}\BibitemShut {NoStop}%
\bibitem [{\citenamefont {Mininni}\ \emph {et~al.}(2003)\citenamefont
  {Mininni}, \citenamefont {G\'omez},\ and\ \citenamefont {Mahajan}}]{Mi2003}%
  \BibitemOpen
  \bibfield  {author} {\bibinfo {author} {\bibfnamefont {P.~D.}\ \bibnamefont
  {Mininni}}, \bibinfo {author} {\bibfnamefont {D.~O.}\ \bibnamefont
  {G\'omez}}, \ and\ \bibinfo {author} {\bibfnamefont {S.~M.}\ \bibnamefont
  {Mahajan}},\ }\href@noop {} {\bibfield  {journal} {\bibinfo  {journal}
  {Astrophys. J.}\ }\textbf {\bibinfo {volume} {584}},\ \bibinfo {pages} {1120}
  (\bibinfo {year} {2003})}\BibitemShut {NoStop}%
\bibitem [{\citenamefont {Matthaeus}\ \emph {et~al.}(2003)\citenamefont
  {Matthaeus}, \citenamefont {Dmitruk}, \citenamefont {Smith}, \citenamefont
  {Ghosh},\ and\ \citenamefont {Oughton}}]{Ma2003}%
  \BibitemOpen
  \bibfield  {author} {\bibinfo {author} {\bibfnamefont {W.~H.}\ \bibnamefont
  {Matthaeus}}, \bibinfo {author} {\bibfnamefont {P.}~\bibnamefont {Dmitruk}},
  \bibinfo {author} {\bibfnamefont {D.}~\bibnamefont {Smith}}, \bibinfo
  {author} {\bibfnamefont {S.}~\bibnamefont {Ghosh}}, \ and\ \bibinfo {author}
  {\bibfnamefont {S.}~\bibnamefont {Oughton}},\ }\href@noop {} {\bibfield
  {journal} {\bibinfo  {journal} {Geophys. Res. Lett.}\ }\textbf {\bibinfo
  {volume} {30}},\ \bibinfo {pages} {2104} (\bibinfo {year}
  {2003})}\BibitemShut {NoStop}%
\bibitem [{\citenamefont {Dmitruk}\ and\ \citenamefont
  {Matthaeus}(2006)}]{D2006}%
  \BibitemOpen
  \bibfield  {author} {\bibinfo {author} {\bibfnamefont {P.}~\bibnamefont
  {Dmitruk}}\ and\ \bibinfo {author} {\bibfnamefont {W.~H.}\ \bibnamefont
  {Matthaeus}},\ }\href@noop {} {\bibfield  {journal} {\bibinfo  {journal}
  {Phys. Plasmas}\ }\textbf {\bibinfo {volume} {13}},\ \bibinfo {pages}
  {042307} (\bibinfo {year} {2006})}\BibitemShut {NoStop}%
\bibitem [{\citenamefont {Galtier}(2008)}]{Ga2008}%
  \BibitemOpen
  \bibfield  {author} {\bibinfo {author} {\bibfnamefont {S.}~\bibnamefont
  {Galtier}},\ }\href@noop {} {\bibfield  {journal} {\bibinfo  {journal} {Phys.
  Rev. E}\ }\textbf {\bibinfo {volume} {77}},\ \bibinfo {pages} {015302}
  (\bibinfo {year} {2008})}\BibitemShut {NoStop}%
\bibitem [{\citenamefont {Kolmogorov}(941b)}]{K1941b}%
  \BibitemOpen
  \bibfield  {author} {\bibinfo {author} {\bibfnamefont {A.~N.}\ \bibnamefont
  {Kolmogorov}},\ }\href@noop {} {\bibfield  {journal} {\bibinfo  {journal}
  {C.R. Acad. Sci.}\ }\textbf {\bibinfo {volume} {32}},\ \bibinfo {pages} {16}
  (\bibinfo {year} {1941b})}\BibitemShut {NoStop}%
\bibitem [{\citenamefont {Weygand}\ \emph {et~al.}(2007)\citenamefont
  {Weygand}, \citenamefont {Matthaeus}, \citenamefont {Dasso}, \citenamefont
  {Kivelson},\ and\ \citenamefont {Walker}}]{WEY2007}%
  \BibitemOpen
  \bibfield  {author} {\bibinfo {author} {\bibfnamefont {J.~M.}\ \bibnamefont
  {Weygand}}, \bibinfo {author} {\bibfnamefont {W.~H.}\ \bibnamefont
  {Matthaeus}}, \bibinfo {author} {\bibfnamefont {S.}~\bibnamefont {Dasso}},
  \bibinfo {author} {\bibfnamefont {M.~G.}\ \bibnamefont {Kivelson}}, \ and\
  \bibinfo {author} {\bibfnamefont {R.~J.}\ \bibnamefont {Walker}},\
  }\href@noop {} {\bibfield  {journal} {\bibinfo  {journal} {J. Geophys. Res.:
  Space Phys.}\ }\textbf {\bibinfo {volume} {112}},\ \bibinfo {pages} {A10}
  (\bibinfo {year} {2007})}\BibitemShut {NoStop}%
\bibitem [{\citenamefont {Mininni}\ and\ \citenamefont
  {Pouquet}(2009)}]{MI2009F}%
  \BibitemOpen
  \bibfield  {author} {\bibinfo {author} {\bibfnamefont {P.~D.}\ \bibnamefont
  {Mininni}}\ and\ \bibinfo {author} {\bibfnamefont {A.}~\bibnamefont
  {Pouquet}},\ }\href@noop {} {\bibfield  {journal} {\bibinfo  {journal} {Phys.
  Rev. E}\ }\textbf {\bibinfo {volume} {80}},\ \bibinfo {pages} {025401}
  (\bibinfo {year} {2009})}\BibitemShut {NoStop}%
\bibitem [{\citenamefont {Boldyrev}\ \emph {et~al.}(2009)\citenamefont
  {Boldyrev}, \citenamefont {Mason},\ and\ \citenamefont {Cattaneo}}]{Bo2009}%
  \BibitemOpen
  \bibfield  {author} {\bibinfo {author} {\bibfnamefont {S.}~\bibnamefont
  {Boldyrev}}, \bibinfo {author} {\bibfnamefont {J.}~\bibnamefont {Mason}}, \
  and\ \bibinfo {author} {\bibfnamefont {F.}~\bibnamefont {Cattaneo}},\
  }\href@noop {} {\bibfield  {journal} {\bibinfo  {journal} {Astrophys. J.
  Letters}\ }\textbf {\bibinfo {volume} {699}},\ \bibinfo {pages} {L39}
  (\bibinfo {year} {2009})}\BibitemShut {NoStop}%
\bibitem [{\citenamefont {Wan}\ \emph {et~al.}(2010)\citenamefont {Wan},
  \citenamefont {Servidio}, \citenamefont {Oughton},\ and\ \citenamefont
  {Matthaeus}}]{W2010}%
  \BibitemOpen
  \bibfield  {author} {\bibinfo {author} {\bibfnamefont {M.}~\bibnamefont
  {Wan}}, \bibinfo {author} {\bibfnamefont {S.}~\bibnamefont {Servidio}},
  \bibinfo {author} {\bibfnamefont {S.}~\bibnamefont {Oughton}}, \ and\
  \bibinfo {author} {\bibfnamefont {W.~H.}\ \bibnamefont {Matthaeus}},\
  }\href@noop {} {\bibfield  {journal} {\bibinfo  {journal} {Phys. Plasmas}\
  }\textbf {\bibinfo {volume} {17}},\ \bibinfo {pages} {052307} (\bibinfo
  {year} {2010})}\BibitemShut {NoStop}%
\bibitem [{\citenamefont {Yoshimatsu}(2012)}]{Y2012}%
  \BibitemOpen
  \bibfield  {author} {\bibinfo {author} {\bibfnamefont {K.}~\bibnamefont
  {Yoshimatsu}},\ }\href@noop {} {\bibfield  {journal} {\bibinfo  {journal}
  {Phys. Rev. E}\ }\textbf {\bibinfo {volume} {85}},\ \bibinfo {pages} {066313}
  (\bibinfo {year} {2012})}\BibitemShut {NoStop}%
\bibitem [{\citenamefont {Sorriso-Valvo}\ \emph {et~al.}(2007)\citenamefont
  {Sorriso-Valvo}, \citenamefont {Marino}, \citenamefont {Carbone},
  \citenamefont {Noullez}, \citenamefont {Lepreti}, \citenamefont {Veltri},
  \citenamefont {Bruno}, \citenamefont {Bavassano},\ and\ \citenamefont
  {Pietropaolo}}]{SV2007}%
  \BibitemOpen
  \bibfield  {author} {\bibinfo {author} {\bibfnamefont {L.}~\bibnamefont
  {Sorriso-Valvo}}, \bibinfo {author} {\bibfnamefont {R.}~\bibnamefont
  {Marino}}, \bibinfo {author} {\bibfnamefont {V.}~\bibnamefont {Carbone}},
  \bibinfo {author} {\bibfnamefont {A.}~\bibnamefont {Noullez}}, \bibinfo
  {author} {\bibfnamefont {F.}~\bibnamefont {Lepreti}}, \bibinfo {author}
  {\bibfnamefont {P.}~\bibnamefont {Veltri}}, \bibinfo {author} {\bibfnamefont
  {R.}~\bibnamefont {Bruno}}, \bibinfo {author} {\bibfnamefont
  {B.}~\bibnamefont {Bavassano}}, \ and\ \bibinfo {author} {\bibfnamefont
  {E.}~\bibnamefont {Pietropaolo}},\ }\href@noop {} {\bibfield  {journal}
  {\bibinfo  {journal} {Phys. Rev. Lett.}\ }\textbf {\bibinfo {volume} {99}},\
  \bibinfo {pages} {115001} (\bibinfo {year} {2007})}\BibitemShut {NoStop}%
\bibitem [{\citenamefont {Sahraoui}(2008)}]{Sa2008}%
  \BibitemOpen
  \bibfield  {author} {\bibinfo {author} {\bibfnamefont {F.}~\bibnamefont
  {Sahraoui}},\ }\href@noop {} {\bibfield  {journal} {\bibinfo  {journal}
  {Phys. Rev. E}\ }\textbf {\bibinfo {volume} {78}},\ \bibinfo {pages} {026402}
  (\bibinfo {year} {2008})}\BibitemShut {NoStop}%
\bibitem [{\citenamefont {Marino}\ \emph {et~al.}(2008)\citenamefont {Marino},
  \citenamefont {Sorriso-Valvo}, \citenamefont {Carbone}, \citenamefont
  {Noullez}, \citenamefont {Bruno},\ and\ \citenamefont {Bavassano}}]{M2008}%
  \BibitemOpen
  \bibfield  {author} {\bibinfo {author} {\bibfnamefont {R.}~\bibnamefont
  {Marino}}, \bibinfo {author} {\bibfnamefont {L.}~\bibnamefont
  {Sorriso-Valvo}}, \bibinfo {author} {\bibfnamefont {V.}~\bibnamefont
  {Carbone}}, \bibinfo {author} {\bibfnamefont {A.}~\bibnamefont {Noullez}},
  \bibinfo {author} {\bibfnamefont {R.}~\bibnamefont {Bruno}}, \ and\ \bibinfo
  {author} {\bibfnamefont {B.}~\bibnamefont {Bavassano}},\ }\href@noop {}
  {\bibfield  {journal} {\bibinfo  {journal} {Astrophys. J. Lett.}\ }\textbf
  {\bibinfo {volume} {677}} (\bibinfo {year} {2008})}\BibitemShut {NoStop}%
\bibitem [{\citenamefont {Matthaeus}\ \emph {et~al.}(1999)\citenamefont
  {Matthaeus}, \citenamefont {Zank}, \citenamefont {Smith},\ and\ \citenamefont
  {Oughton}}]{M1999}%
  \BibitemOpen
  \bibfield  {author} {\bibinfo {author} {\bibfnamefont {W.~H.}\ \bibnamefont
  {Matthaeus}}, \bibinfo {author} {\bibfnamefont {G.~P.}\ \bibnamefont {Zank}},
  \bibinfo {author} {\bibfnamefont {C.~W.}\ \bibnamefont {Smith}}, \ and\
  \bibinfo {author} {\bibfnamefont {S.}~\bibnamefont {Oughton}},\ }\href@noop
  {} {\bibfield  {journal} {\bibinfo  {journal} {Phys. Rev. Lett.}\ }\textbf
  {\bibinfo {volume} {82}},\ \bibinfo {pages} {3444} (\bibinfo {year}
  {1999})}\BibitemShut {NoStop}%
\bibitem [{\citenamefont {MacBride}\ \emph {et~al.}(2008)\citenamefont
  {MacBride}, \citenamefont {Smith},\ and\ \citenamefont {Forman}}]{MB2008}%
  \BibitemOpen
  \bibfield  {author} {\bibinfo {author} {\bibfnamefont {B.~T.}\ \bibnamefont
  {MacBride}}, \bibinfo {author} {\bibfnamefont {C.~W.}\ \bibnamefont {Smith}},
  \ and\ \bibinfo {author} {\bibfnamefont {F.~A.}\ \bibnamefont {Forman}},\
  }\href@noop {} {\bibfield  {journal} {\bibinfo  {journal} {ApJ}\ }\textbf
  {\bibinfo {volume} {679}},\ \bibinfo {pages} {1644} (\bibinfo {year}
  {2008})}\BibitemShut {NoStop}%
\bibitem [{\citenamefont {Davidson}(2009)}]{Da2009}%
  \BibitemOpen
  \bibfield  {author} {\bibinfo {author} {\bibfnamefont {P.~A.}\ \bibnamefont
  {Davidson}},\ }\href@noop {} {\bibfield  {journal} {\bibinfo  {journal} {J.
  Fluid Mech.}\ }\textbf {\bibinfo {volume} {632}},\ \bibinfo {pages} {329}
  (\bibinfo {year} {2009})}\BibitemShut {NoStop}%
\bibitem [{\citenamefont {Wan}\ \emph {et~al.}(2012)\citenamefont {Wan},
  \citenamefont {Oughton}, \citenamefont {Servidio},\ and\ \citenamefont
  {Matthaeus}}]{W2012}%
  \BibitemOpen
  \bibfield  {author} {\bibinfo {author} {\bibfnamefont {M.}~\bibnamefont
  {Wan}}, \bibinfo {author} {\bibfnamefont {S.}~\bibnamefont {Oughton}},
  \bibinfo {author} {\bibfnamefont {S.}~\bibnamefont {Servidio}}, \ and\
  \bibinfo {author} {\bibfnamefont {W.~H.}\ \bibnamefont {Matthaeus}},\
  }\href@noop {} {\bibfield  {journal} {\bibinfo  {journal} {J. Fluid Mech.}\
  }\textbf {\bibinfo {volume} {697}},\ \bibinfo {pages} {296} (\bibinfo {year}
  {2012})}\BibitemShut {NoStop}%
\bibitem [{\citenamefont {Benzi}\ \emph {et~al.}(1993)\citenamefont {Benzi},
  \citenamefont {Ciliberto}, \citenamefont {Tripiccione}, \citenamefont
  {Baudet}, \citenamefont {Massaioli},\ and\ \citenamefont {Succi}}]{R1993}%
  \BibitemOpen
  \bibfield  {author} {\bibinfo {author} {\bibfnamefont {R.}~\bibnamefont
  {Benzi}}, \bibinfo {author} {\bibfnamefont {S.}~\bibnamefont {Ciliberto}},
  \bibinfo {author} {\bibfnamefont {R.}~\bibnamefont {Tripiccione}}, \bibinfo
  {author} {\bibfnamefont {C.}~\bibnamefont {Baudet}}, \bibinfo {author}
  {\bibfnamefont {F.}~\bibnamefont {Massaioli}}, \ and\ \bibinfo {author}
  {\bibfnamefont {S.}~\bibnamefont {Succi}},\ }\href@noop {} {\bibfield
  {journal} {\bibinfo  {journal} {Phys. Rev. E}\ }\textbf {\bibinfo {volume}
  {48}},\ \bibinfo {pages} {R29} (\bibinfo {year} {1993})}\BibitemShut
  {NoStop}%
\bibitem [{\citenamefont {Grossmann}\ \emph {et~al.}(1997)\citenamefont
  {Grossmann}, \citenamefont {Lohse},\ and\ \citenamefont {Reeh}}]{G1997}%
  \BibitemOpen
  \bibfield  {author} {\bibinfo {author} {\bibfnamefont {S.}~\bibnamefont
  {Grossmann}}, \bibinfo {author} {\bibfnamefont {D.}~\bibnamefont {Lohse}}, \
  and\ \bibinfo {author} {\bibfnamefont {A.}~\bibnamefont {Reeh}},\ }\href@noop
  {} {\bibfield  {journal} {\bibinfo  {journal} {Phys. Rev. E}\ }\textbf
  {\bibinfo {volume} {56}},\ \bibinfo {pages} {5473} (\bibinfo {year}
  {1997})}\BibitemShut {NoStop}%
\bibitem [{\citenamefont {Matthaeus}\ and\ \citenamefont
  {Velli}(2011)}]{M2011}%
  \BibitemOpen
  \bibfield  {author} {\bibinfo {author} {\bibfnamefont {W.}~\bibnamefont
  {Matthaeus}}\ and\ \bibinfo {author} {\bibfnamefont {M.}~\bibnamefont
  {Velli}},\ }\href@noop {} {\bibfield  {journal} {\bibinfo  {journal} {Space
  Science Reviews}\ }\textbf {\bibinfo {volume} {160}},\ \bibinfo {pages} {145}
  (\bibinfo {year} {2011})}\BibitemShut {NoStop}%
\bibitem [{\citenamefont {Oughton}\ \emph {et~al.}(1997)\citenamefont
  {Oughton}, \citenamefont {Radler},\ and\ \citenamefont
  {Matthaeus}}]{ORM1997}%
  \BibitemOpen
  \bibfield  {author} {\bibinfo {author} {\bibfnamefont {S.}~\bibnamefont
  {Oughton}}, \bibinfo {author} {\bibfnamefont {K.~H.}\ \bibnamefont {Radler}},
  \ and\ \bibinfo {author} {\bibfnamefont {W.~H.}\ \bibnamefont {Matthaeus}},\
  }\href@noop {} {\bibfield  {journal} {\bibinfo  {journal} {Phys. Rev. E}\
  }\textbf {\bibinfo {volume} {56}},\ \bibinfo {pages} {2875} (\bibinfo {year}
  {1997})}\BibitemShut {NoStop}%
\bibitem [{\citenamefont {Batchelor}(1953)}]{Ba1953}%
  \BibitemOpen
  \bibfield  {author} {\bibinfo {author} {\bibfnamefont {G.~K.}\ \bibnamefont
  {Batchelor}},\ }\href@noop {} {\emph {\bibinfo {title} {The theory of
  homogeneus turbulence}}}\ (\bibinfo  {publisher} {Cambridge Univ. Press},\
  \bibinfo {year} {1953})\BibitemShut {NoStop}%
\bibitem [{\citenamefont {Chkhetiani}(1996)}]{Ch1996}%
  \BibitemOpen
  \bibfield  {author} {\bibinfo {author} {\bibfnamefont {O.}~\bibnamefont
  {Chkhetiani}},\ }\href@noop {} {\bibfield  {journal} {\bibinfo  {journal}
  {J.E.T.P. Lett.}\ }\textbf {\bibinfo {volume} {10}},\ \bibinfo {pages} {808}
  (\bibinfo {year} {1996})}\BibitemShut {NoStop}%
\bibitem [{\citenamefont {{G{o}mez}}\ \emph {et~al.}(2000)\citenamefont
  {{G{o}mez}}, \citenamefont {Politano},\ and\ \citenamefont
  {Pouquet}}]{G2000}%
  \BibitemOpen
  \bibfield  {author} {\bibinfo {author} {\bibfnamefont {T.}~\bibnamefont
  {{G{o}mez}}}, \bibinfo {author} {\bibfnamefont {H.}~\bibnamefont {Politano}},
  \ and\ \bibinfo {author} {\bibfnamefont {A.}~\bibnamefont {Pouquet}},\
  }\href@noop {} {\bibfield  {journal} {\bibinfo  {journal} {Phys. Rev. E}\
  }\textbf {\bibinfo {volume} {61}},\ \bibinfo {pages} {5321} (\bibinfo {year}
  {2000})}\BibitemShut {NoStop}%
\bibitem [{\citenamefont {Kurien}(2003)}]{K2003}%
  \BibitemOpen
  \bibfield  {author} {\bibinfo {author} {\bibfnamefont {S.}~\bibnamefont
  {Kurien}},\ }\href@noop {} {\bibfield  {journal} {\bibinfo  {journal}
  {Physical D}\ }\textbf {\bibinfo {volume} {175}},\ \bibinfo {pages} {167}
  (\bibinfo {year} {2003})}\BibitemShut {NoStop}%
\bibitem [{\citenamefont {Andr\'es}\ \emph
  {et~al.}(2014{\natexlab{a}})\citenamefont {Andr\'es}, \citenamefont {Martin},
  \citenamefont {Dmitruk},\ and\ \citenamefont {G\'omez}}]{A2014a}%
  \BibitemOpen
  \bibfield  {author} {\bibinfo {author} {\bibfnamefont {N.}~\bibnamefont
  {Andr\'es}}, \bibinfo {author} {\bibfnamefont {L.~N.}\ \bibnamefont
  {Martin}}, \bibinfo {author} {\bibfnamefont {P.}~\bibnamefont {Dmitruk}}, \
  and\ \bibinfo {author} {\bibfnamefont {D.~O.}\ \bibnamefont {G\'omez}},\
  }\href@noop {} {\bibfield  {journal} {\bibinfo  {journal} {Phys. Plasmas}\
  }\textbf {\bibinfo {volume} {21}},\ \bibinfo {pages} {072904} (\bibinfo
  {year} {2014}{\natexlab{a}})}\BibitemShut {NoStop}%
\bibitem [{\citenamefont {Andr\'es}\ \emph
  {et~al.}(2014{\natexlab{b}})\citenamefont {Andr\'es}, \citenamefont
  {Gonzalez}, \citenamefont {Martin}, \citenamefont {Dmitruk},\ and\
  \citenamefont {G\'omez}}]{A2014b}%
  \BibitemOpen
  \bibfield  {author} {\bibinfo {author} {\bibfnamefont {N.}~\bibnamefont
  {Andr\'es}}, \bibinfo {author} {\bibfnamefont {C.}~\bibnamefont {Gonzalez}},
  \bibinfo {author} {\bibfnamefont {L.~N.}\ \bibnamefont {Martin}}, \bibinfo
  {author} {\bibfnamefont {P.}~\bibnamefont {Dmitruk}}, \ and\ \bibinfo
  {author} {\bibfnamefont {D.~O.}\ \bibnamefont {G\'omez}},\ }\href@noop {}
  {\bibfield  {journal} {\bibinfo  {journal} {Phys. Plasmas}\ }\textbf
  {\bibinfo {volume} {21}},\ \bibinfo {pages} {122305} (\bibinfo {year}
  {2014}{\natexlab{b}})}\BibitemShut {NoStop}%
\bibitem [{\citenamefont {Andr\'es}\ \emph {et~al.}(2016)\citenamefont
  {Andr\'es}, \citenamefont {Dmitruk},\ and\ \citenamefont {G\'omez}}]{A2016}%
  \BibitemOpen
  \bibfield  {author} {\bibinfo {author} {\bibfnamefont {N.}~\bibnamefont
  {Andr\'es}}, \bibinfo {author} {\bibfnamefont {P.}~\bibnamefont {Dmitruk}}, \
  and\ \bibinfo {author} {\bibfnamefont {D.~O.}\ \bibnamefont {G\'omez}},\
  }\href@noop {} {\bibfield  {journal} {\bibinfo  {journal} {Phys. Plasmas}\
  }\textbf {\bibinfo {volume} {23}},\ \bibinfo {pages} {022903} (\bibinfo
  {year} {2016})}\BibitemShut {NoStop}%
\bibitem [{\citenamefont {Matthaeus}\ and\ \citenamefont
  {Smith}(1981)}]{M1981}%
  \BibitemOpen
  \bibfield  {author} {\bibinfo {author} {\bibfnamefont {W.~H.}\ \bibnamefont
  {Matthaeus}}\ and\ \bibinfo {author} {\bibfnamefont {C.}~\bibnamefont
  {Smith}},\ }\href@noop {} {\bibfield  {journal} {\bibinfo  {journal} {Phys.
  Rev. A}\ }\textbf {\bibinfo {volume} {24}},\ \bibinfo {pages} {2135}
  (\bibinfo {year} {1981})}\BibitemShut {NoStop}%
\bibitem [{\citenamefont {Podesta}(2008)}]{P2008}%
  \BibitemOpen
  \bibfield  {author} {\bibinfo {author} {\bibfnamefont {J.~J.}\ \bibnamefont
  {Podesta}},\ }\href@noop {} {\bibfield  {journal} {\bibinfo  {journal} {Jour.
  Fluid Mech.}\ }\textbf {\bibinfo {volume} {609}},\ \bibinfo {pages} {171}
  (\bibinfo {year} {2008})}\BibitemShut {NoStop}%
\bibitem [{\citenamefont {Carbone}\ \emph {et~al.}(2009)\citenamefont
  {Carbone}, \citenamefont {Sorriso-Valvo},\ and\ \citenamefont
  {Marino}}]{C2009}%
  \BibitemOpen
  \bibfield  {author} {\bibinfo {author} {\bibfnamefont {V.}~\bibnamefont
  {Carbone}}, \bibinfo {author} {\bibfnamefont {L.}~\bibnamefont
  {Sorriso-Valvo}}, \ and\ \bibinfo {author} {\bibfnamefont {R.}~\bibnamefont
  {Marino}},\ }\href@noop {} {\bibfield  {journal} {\bibinfo  {journal} {EPL
  (Europhysics Letters)}\ }\textbf {\bibinfo {volume} {88}},\ \bibinfo {pages}
  {25001} (\bibinfo {year} {2009})}\BibitemShut {NoStop}%
\bibitem [{\citenamefont {Sahraoui}\ \emph {et~al.}(2009)\citenamefont
  {Sahraoui}, \citenamefont {Goldstein}, \citenamefont {Robert},\ and\
  \citenamefont {Khotyaintsev}}]{S2009}%
  \BibitemOpen
  \bibfield  {author} {\bibinfo {author} {\bibfnamefont {F.}~\bibnamefont
  {Sahraoui}}, \bibinfo {author} {\bibfnamefont {M.~L.}\ \bibnamefont
  {Goldstein}}, \bibinfo {author} {\bibfnamefont {P.}~\bibnamefont {Robert}}, \
  and\ \bibinfo {author} {\bibfnamefont {Y.~V.}\ \bibnamefont {Khotyaintsev}},\
  }\href@noop {} {\bibfield  {journal} {\bibinfo  {journal} {Phys. Rev. Lett.}\
  }\textbf {\bibinfo {volume} {102}},\ \bibinfo {pages} {231102} (\bibinfo
  {year} {2009})}\BibitemShut {NoStop}%
\bibitem [{\citenamefont {Martin}\ \emph {et~al.}(2010)\citenamefont {Martin},
  \citenamefont {Dmitruk},\ and\ \citenamefont {G\'omez}}]{M2010}%
  \BibitemOpen
  \bibfield  {author} {\bibinfo {author} {\bibfnamefont {L.~N.}\ \bibnamefont
  {Martin}}, \bibinfo {author} {\bibfnamefont {P.}~\bibnamefont {Dmitruk}}, \
  and\ \bibinfo {author} {\bibfnamefont {D.~O.}\ \bibnamefont {G\'omez}},\
  }\href@noop {} {\bibfield  {journal} {\bibinfo  {journal} {Phys. Plasmas}\
  }\textbf {\bibinfo {volume} {17}},\ \bibinfo {pages} {112304} (\bibinfo
  {year} {2010})}\BibitemShut {NoStop}%
\bibitem [{\citenamefont {{G{\'o}mez}}\ \emph {et~al.}(2008)\citenamefont
  {{G{\'o}mez}}, \citenamefont {{Mahajan}},\ and\ \citenamefont
  {{Dmitruk}}}]{G2008}%
  \BibitemOpen
  \bibfield  {author} {\bibinfo {author} {\bibfnamefont {D.~O.}\ \bibnamefont
  {{G{\'o}mez}}}, \bibinfo {author} {\bibfnamefont {S.~M.}\ \bibnamefont
  {{Mahajan}}}, \ and\ \bibinfo {author} {\bibfnamefont {P.}~\bibnamefont
  {{Dmitruk}}},\ }\href@noop {} {\bibfield  {journal} {\bibinfo  {journal}
  {Phys. Plasmas}\ }\textbf {\bibinfo {volume} {15}},\ \bibinfo {pages}
  {102303} (\bibinfo {year} {2008})}\BibitemShut {NoStop}%
\bibitem [{\citenamefont {Mininni}\ \emph {et~al.}(2007)\citenamefont
  {Mininni}, \citenamefont {Alekaxis},\ and\ \citenamefont {Pouquet}}]{Mi2007}%
  \BibitemOpen
  \bibfield  {author} {\bibinfo {author} {\bibfnamefont {P.~D.}\ \bibnamefont
  {Mininni}}, \bibinfo {author} {\bibfnamefont {A.}~\bibnamefont {Alekaxis}}, \
  and\ \bibinfo {author} {\bibfnamefont {A.}~\bibnamefont {Pouquet}},\
  }\href@noop {} {\bibfield  {journal} {\bibinfo  {journal} {J. Plasma Phys.}\
  }\textbf {\bibinfo {volume} {73}},\ \bibinfo {pages} {377} (\bibinfo {year}
  {2007})}\BibitemShut {NoStop}%
\bibitem [{\citenamefont {Biskamp}\ \emph {et~al.}(1997)\citenamefont
  {Biskamp}, \citenamefont {Schwarz},\ and\ \citenamefont {Drake}}]{Bi1997}%
  \BibitemOpen
  \bibfield  {author} {\bibinfo {author} {\bibfnamefont {D.}~\bibnamefont
  {Biskamp}}, \bibinfo {author} {\bibfnamefont {E.}~\bibnamefont {Schwarz}}, \
  and\ \bibinfo {author} {\bibfnamefont {J.~F.}\ \bibnamefont {Drake}},\
  }\href@noop {} {\bibfield  {journal} {\bibinfo  {journal} {Physics of
  Plasmas}\ }\textbf {\bibinfo {volume} {4}},\ \bibinfo {pages} {1002}
  (\bibinfo {year} {1997})}\BibitemShut {NoStop}%
\bibitem [{\citenamefont {Meyrand}\ and\ \citenamefont
  {Galtier}(2010)}]{MG2010}%
  \BibitemOpen
  \bibfield  {author} {\bibinfo {author} {\bibfnamefont {R.}~\bibnamefont
  {Meyrand}}\ and\ \bibinfo {author} {\bibfnamefont {S.}~\bibnamefont
  {Galtier}},\ }\href@noop {} {\bibfield  {journal} {\bibinfo  {journal} {The
  Astrophysical Journal}\ }\textbf {\bibinfo {volume} {721}},\ \bibinfo {pages}
  {1421} (\bibinfo {year} {2010})}\BibitemShut {NoStop}%
\bibitem [{\citenamefont {Sahraoui}\ \emph {et~al.}(2010)\citenamefont
  {Sahraoui}, \citenamefont {Goldstein}, \citenamefont {Belmont}, \citenamefont
  {Canu},\ and\ \citenamefont {Rezeau}}]{S2010}%
  \BibitemOpen
  \bibfield  {author} {\bibinfo {author} {\bibfnamefont {F.}~\bibnamefont
  {Sahraoui}}, \bibinfo {author} {\bibfnamefont {M.~L.}\ \bibnamefont
  {Goldstein}}, \bibinfo {author} {\bibfnamefont {G.}~\bibnamefont {Belmont}},
  \bibinfo {author} {\bibfnamefont {P.}~\bibnamefont {Canu}}, \ and\ \bibinfo
  {author} {\bibfnamefont {L.}~\bibnamefont {Rezeau}},\ }\href@noop {}
  {\bibfield  {journal} {\bibinfo  {journal} {Phys. Rev. Lett.}\ }\textbf
  {\bibinfo {volume} {105}},\ \bibinfo {pages} {131101} (\bibinfo {year}
  {2010})}\BibitemShut {NoStop}%
\bibitem [{\citenamefont {Sahraoui}\ \emph {et~al.}(2011)\citenamefont
  {Sahraoui}, \citenamefont {Goldstein}, \citenamefont {Abdul-Kader},
  \citenamefont {Belmont}, \citenamefont {Rezeau},\ and\ \citenamefont
  {Robert}}]{S2011}%
  \BibitemOpen
  \bibfield  {author} {\bibinfo {author} {\bibfnamefont {F.}~\bibnamefont
  {Sahraoui}}, \bibinfo {author} {\bibfnamefont {M.~L.}\ \bibnamefont
  {Goldstein}}, \bibinfo {author} {\bibfnamefont {K.}~\bibnamefont
  {Abdul-Kader}}, \bibinfo {author} {\bibfnamefont {G.}~\bibnamefont
  {Belmont}}, \bibinfo {author} {\bibfnamefont {L.}~\bibnamefont {Rezeau}}, \
  and\ \bibinfo {author} {\bibfnamefont {P.}~\bibnamefont {Robert}},\
  }\href@noop {} {\bibfield  {journal} {\bibinfo  {journal} {C. R. Physique}\
  }\textbf {\bibinfo {volume} {12}},\ \bibinfo {pages} {132} (\bibinfo {year}
  {2011})}\BibitemShut {NoStop}%
\bibitem [{\citenamefont {Alexandrova}\ \emph {et~al.}(2009)\citenamefont
  {Alexandrova}, \citenamefont {Saur}, \citenamefont {Lacombe}, \citenamefont
  {Mangeney}, \citenamefont {Mitchell}, \citenamefont {Schwartz},\ and\
  \citenamefont {Robert}}]{A2009}%
  \BibitemOpen
  \bibfield  {author} {\bibinfo {author} {\bibfnamefont {O.}~\bibnamefont
  {Alexandrova}}, \bibinfo {author} {\bibfnamefont {J.}~\bibnamefont {Saur}},
  \bibinfo {author} {\bibfnamefont {C.}~\bibnamefont {Lacombe}}, \bibinfo
  {author} {\bibfnamefont {A.}~\bibnamefont {Mangeney}}, \bibinfo {author}
  {\bibfnamefont {J.}~\bibnamefont {Mitchell}}, \bibinfo {author}
  {\bibfnamefont {S.~J.}\ \bibnamefont {Schwartz}}, \ and\ \bibinfo {author}
  {\bibfnamefont {P.}~\bibnamefont {Robert}},\ }\href@noop {} {\bibfield
  {journal} {\bibinfo  {journal} {Phys. Rev. Lett.}\ }\textbf {\bibinfo
  {volume} {103}},\ \bibinfo {pages} {165003} (\bibinfo {year}
  {2009})}\BibitemShut {NoStop}%
\bibitem [{\citenamefont {{Pietarila Graham}}\ \emph
  {et~al.}(2006)\citenamefont {{Pietarila Graham}}, \citenamefont {{Holm}},
  \citenamefont {{Mininni}},\ and\ \citenamefont {{Pouquet}}}]{P2006}%
  \BibitemOpen
  \bibfield  {author} {\bibinfo {author} {\bibfnamefont {J.}~\bibnamefont
  {{Pietarila Graham}}}, \bibinfo {author} {\bibfnamefont {D.~D.}\ \bibnamefont
  {{Holm}}}, \bibinfo {author} {\bibfnamefont {P.}~\bibnamefont {{Mininni}}}, \
  and\ \bibinfo {author} {\bibfnamefont {A.}~\bibnamefont {{Pouquet}}},\
  }\href@noop {} {\bibfield  {journal} {\bibinfo  {journal} {Phys. Fluids}\
  }\textbf {\bibinfo {volume} {18}},\ \bibinfo {pages} {045106} (\bibinfo
  {year} {2006})}\BibitemShut {NoStop}%
\bibitem [{\citenamefont {Holm}(2002)}]{H2002}%
  \BibitemOpen
  \bibfield  {author} {\bibinfo {author} {\bibfnamefont {D.~D.}\ \bibnamefont
  {Holm}},\ }\href@noop {} {\bibfield  {journal} {\bibinfo  {journal} {J. Fluid
  Mech.}\ }\textbf {\bibinfo {volume} {467}},\ \bibinfo {pages} {205} (\bibinfo
  {year} {2002})}\BibitemShut {NoStop}%
\bibitem [{\citenamefont {Leamon}\ \emph {et~al.}(1998)\citenamefont {Leamon},
  \citenamefont {Matthaeus}, \citenamefont {Smith},\ and\ \citenamefont
  {K.}}]{L1998}%
  \BibitemOpen
  \bibfield  {author} {\bibinfo {author} {\bibfnamefont {R.~J.}\ \bibnamefont
  {Leamon}}, \bibinfo {author} {\bibfnamefont {W.~H.}\ \bibnamefont
  {Matthaeus}}, \bibinfo {author} {\bibfnamefont {C.~W.}\ \bibnamefont
  {Smith}}, \ and\ \bibinfo {author} {\bibfnamefont {W.~H.}\ \bibnamefont
  {K.}},\ }\href@noop {} {\bibfield  {journal} {\bibinfo  {journal} {Astrophys.
  J.}\ }\textbf {\bibinfo {volume} {507}},\ \bibinfo {pages} {L181} (\bibinfo
  {year} {1998})}\BibitemShut {NoStop}%
\end{thebibliography}%
\end{document}